\documentclass[journal,10pt,draftclsnofoot,onecolumn]{IEEEtran}
\usepackage[latin1]{inputenc}
\usepackage{amssymb}
\usepackage[dvips,pdftex]{graphicx}
\usepackage{amssymb}
\usepackage{caption}
\usepackage{subcaption}

\usepackage{verbatim}
\usepackage[normalem]{ulem}
\usepackage{lineno}
\usepackage{setspace}
\usepackage{color}
\usepackage{url}
\usepackage{cite}

\usepackage{float}
\usepackage{xspace}
\usepackage{multirow}

\usepackage{amsmath}

\usepackage{supertabular}

\IEEEoverridecommandlockouts

\renewcommand{\textbf}[1]{\begingroup\bfseries\mathversion{bold}#1\endgroup}

\begin{document}

\title{Rethinking the Contention Resolution Mechanism in WiMAX Networks using Lattice Correlators for Improved Smart Grid Communication Performance
}

\author{Martin L\'evesque

\thanks{Corresponding author: Martin L\'evesque (email: \texttt{levesque.martin@gmail.com}).}
}


\IEEEaftertitletext{\vspace*{-.7cm}
}

\maketitle
\begin{abstract}
By using experimental measurements of smart grid applications compliant with IEC 61850 in trace-driven WiMAX simulations, we show that the WiMAX MAC protocol efficiency decreases as a function of the number of stations. To avoid this shortcoming, we propose and analyze a novel WiMAX MAC protocol for smart grid applications, which uses lattice correlators to improve the throughput-delay performance significantly. For the considered configurations, the obtained maximum throughput of the proposed MAC protocol outperforms the current WiMAX MAC protocol by up to 41\%.
\end{abstract}
\vspace{-.25cm}

\section{Introduction}
Research on a new power system paradigm, also known as smart grid, has recently attracted significant attention in the information technology, telecommunications, and power system communities \cite{6157575}. WiMAX is particularly interesting for smart grid communications, especially in Canada where a separate spectrum (30 MHz in the 1.8 GHz frequency band) is dedicated to utilities, and also provides long-range connectivity \cite{lteSG}. 

As we show in the next section, however, WiMAX is not efficient as the number of stations increases, mainly due to the contention resolution mechanism in use. Rethinking wireless medium access control (MAC) protocols was the object of many recent research studies on IEEE 802.11 wireless local area networks (WLANs), exploiting the frequency domain (rather than time domain) for contention resolution \cite{6195796}. In particular, the proposed WiFi-Nano solution aims at rethinking WiFi by significantly reducing the contention resolution air time using the novel lattice correlation technique to detect other stations transmitting \cite{wifiNano}. The authors experimentally demonstrated that it is not only possible to detect other stations transmitting, but also to keep the station at the lowest backoff value by sending and sensing so-called speculative preambles. 

The remainder of the letter is structured as follows. In Section \ref{sec:motivation}, we motivate our work by showing experimental traffic traces injected in WiMAX simulations. In Section \ref{sec:proposedProtocol}, we describe our proposed MAC protocol and analyze its performance in Section \ref{sec:analysis}. Section \ref{sec:numericalResults} presents numerical results and Section \ref{sec:conclusions} draws some conclusions.

\section{Motivation}
\label{sec:motivation}
\subsection{Smart Grid Communications - Traffic Characteristics}
An important smart grid application is the monitoring and sensing of a variety of distribution nodes, including substations, high-voltage/low-voltage transformers, and customers. The message payload lengths can significantly affect the overall network performance in terms of throughput and delay. We recorded packets of a traffic simulator based on IEC 61850 and an upcoming renewable energy source web service application developed in our smart grid lab. Note that IEC 61850 is considered to become the de-facto smart grid message standard.

\begin{table}
\caption{Experimental measurements of smart grid applications based on the IEC 61850 standard.}
\label{table:messageConfigurations}
\begin{center}
{\renewcommand{\arraystretch}{1.2}
    \begin{tabular}{ | l | l |}
    \hline
    Source node & Average payload length \\ \hline
	HVA/LV & 500 bytes \\\hline
	Substation & 5000 bytes \\\hline
	DER & 224 bytes \\\hline
	Switch & 100 bytes \\\hline
    \end{tabular}
}
\end{center}
\end{table}

The average payload length originating from high-voltage/low-voltage (HVA/LV) transformers, substations, distributed energy resources (DERs), and controllable switches were found to be 500, 5000, 224, and 100 bytes long, respectively (Table \ref{table:messageConfigurations}). A single variable-value pair accounts for 100 bytes. The measured payload length of 500 bytes for the HVA/LV nodes corresponds to active/reactive power, voltage, current, and position. Except for the substation packets, whose number is limited, all packets have a payload length smaller than or equal to 500 bytes.

\subsection{Performance of WiFi and WiMAX MAC Protocols}
First, the performance of the IEEE 802.11n distributed coordination function (DCF) and WiMAX MAC protocols are evaluated by using the aforementioned measured smart grid payload traces. The focus of this letter is on WiMAX, but WiFi results are also presented for comparison since WiFi could also be used for smart grid communications. In the following, the achieved throughput is measured for a total offered load reaching the upstream capacity, i.e., 300 and 23.5 Mbps for IEEE 802.11n and WiMAX, respectively. Each station sends to the base station (BS). OMNeT++ is used based with parameters of both technologies set to the values used in \cite{wcncWiFi} for WiFi and \cite{wimaxPerfEval} for WiMAX for best-effort service.

\begin{figure}
\centering
\includegraphics[width=.50\textwidth]{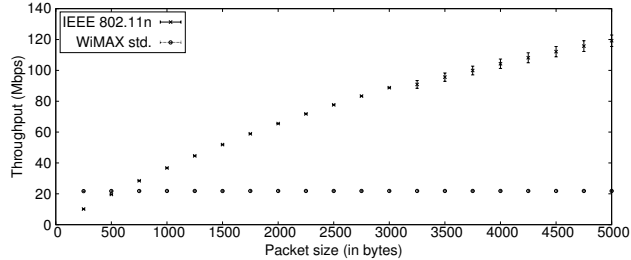}
\caption{Throughput performance for different payload sizes (number of stations: 10, confidence interval (CI): 95\%).}
\label{fig:motivationPktLen}
\end{figure}

The payload length may have a significant impact on the communications efficiency. Therefore, the payload length is varied with 10 stations sending to a given access point, as depicted in Fig. \ref{fig:motivationPktLen}. For IEEE 802.11n, the achieved throughput is significantly influenced by the packet length, which is widely known due to signaling and collision resolution overheads \cite{wifiNano}. For WiMAX with best-effort service, the achieved throughput does not vary much, depending on the payload length for a small number of transmitting nodes since each station can send multiple frames stored its queue by using the reserved subchannel(s). However, smart grid nodes should not have to queue multiple measurement packets, but instead send them as they arrive in their transmission queues without any delay. Note that smart grid applications may involve a large number of nodes transmitting typically at low data rates.
\begin{figure}
\centering
\includegraphics[width=.50\textwidth]{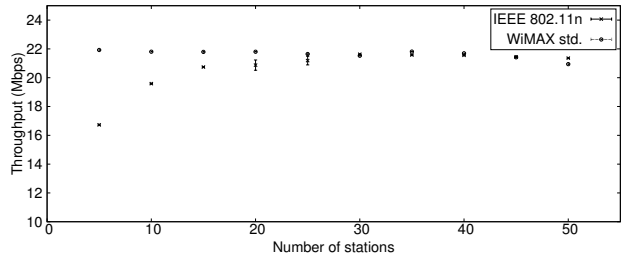}
\caption{Throughput performance for different number of stations (payload length: 500 bytes, CI: 95\%).}
\label{fig:motivationNbNodes}
\end{figure}

\begin{table}
\caption{Obtained WiMAX throughput (th.) with high number of stations (CI: 95 \%).}
\label{table:resultsWiMAXNbNodes}
\begin{center}
{\renewcommand{\arraystretch}{1.2}
    \begin{tabular}{ | l | l | l | l |}
    \hline
    \# nodes & Min. Th. & Mean Th. & Max Th.  \\ \hline
	50 & 20.90 Mbps & 20.94 Mbps & 20.99 Mbps \\\hline
	333 & 15.01 Mbps & 15.05 Mbps & 15.09 Mbps \\\hline
	666 & 13.93 Mbps & 13.95 Mbps & 13.98 Mbps \\\hline
	1000 & 13.74 Mbps & 13.76 Mbps & 13.77 Mbps \\\hline
    \end{tabular}
}
\end{center}
\end{table}

Next, the number of stations is varied to show the throughput performance for a fixed payload length of 500 bytes. A payload length of 500 bytes is used since the quintet of ``active/reactive power, voltage, current, and position" corresponds to the main performance metrics of power systems. Fig. \ref{fig:motivationNbNodes} depicts the throughput performance with different number of stations for both WiFi and WiMAX. With the DCF mechanism used by WiFi, the throughput increases from 5 to 35 stations followed by a drop. However, with WiMAX there is a constant throughput decrease as a function of the number of stations, since the contention period becomes significant and too many stations' transmissions suffer from collision during this period. Furthermore, when the packet size does not exactly equal a certain number of subchannels, a significant amount of unused space is wasted. For example, if the payload length is 500 bytes and the subchannel size is equal to 420 bytes, excluding overheads, 2 reserved subchannels waste $420 \cdot 2 - 500 = 340$ bytes. We also simulated a WiMAX network with a higher number of stations. We obtained a maximum throughput of 15, 13.93, and 13.74 Mbps for 333, 666, and 1000 nodes (Table \ref{table:resultsWiMAXNbNodes}), respectively, and an efficiency of $\frac{13.74\ Mbps}{23.5\ Mbps} = 55\%$ is obtained for the case of 1000 nodes.

\subsection{Problem Statement}
Smart grid sensor communications is characterized by the exchange of periodic short measurements with an average payload length smaller than or equal to 500 bytes, as observed from our experimental traffic measurements of smart grid applications based on IEC standard 61850. By injecting this type of traffic in WiMAX simulations, the overall network efficiency is found to significantly decrease as the number of nodes increases, meaning that a WiMAX channel dedicated to smart grid applications would not be used efficiently. 

\section{Proposed MAC Protocol}
\label{sec:proposedProtocol}
\begin{figure}
\centering
\includegraphics[width=.50\textwidth]{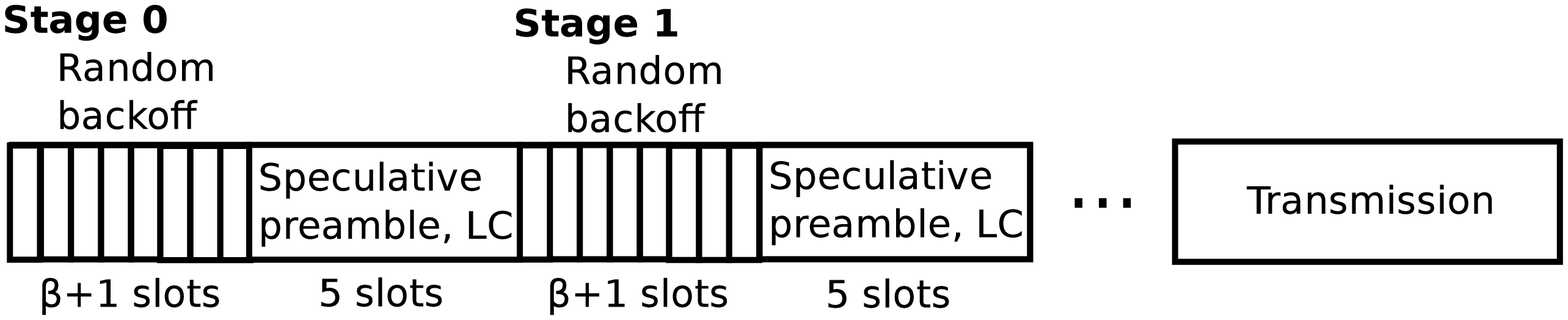}
\caption{Proposed MAC protocol exploiting lattice correlators: At a given stage $s$, competing stations with the same minimum backoff number go to stage $s+1$; when a single station has the minimum backoff number, it wins channel access and starts transmitting.}
\label{fig:proposedMAC}
\end{figure}
In this section, we present our proposed wireless MAC protocol, which exploits lattice correlators to improve network efficiency. The principle is illustrated in Fig. \ref{fig:proposedMAC} and works as follows. At the beginning of an upstream frame, each station having at least one frame in its queue randomly selects a channel in its list of allowed channels allocated by the BS. As shown in Fig. \ref{fig:proposedMAC}, on a given channel $c$, a given station starts competing with other stations that have also selected channel $c$. At stage 0, the station generates a random backoff number $r$ in $[0..\beta]$ and waits for $r \cdot \epsilon$, whereby $\epsilon$ denotes the slot duration. Subsequently, it starts transmitting a speculative preamble (consisting of 5 slots), as described in\cite{wifiNano}, and listens to other stations' preambles. At the end of stage 0, the station proceeds as follows:
\begin{itemize}
\item \textit{Case 1:} It selects the minimum random backoff number $r$ and no other station selected $r$, and thus wins channel access and can start transmitting.
\item \textit{Case 2:} Another station selects a smaller backoff number than $r$, and hence it loses access to channel $c$ and waits for the next upstream frame to compete again.
\item \textit{Case 3:} It selects the minimum random backoff number $r$ and there is one or more stations which selected $r$ as well and thus enters stage 1.
\end{itemize}  
If the third case applies and stage 1 starts, the station again competes in a similar way, but with a lower number of stations, until case 1 holds (up to the maximum stage number $s_{max}$). In doing so, the protocol allows to quickly solve contentions in a distributed manner. 
	
Next, we introduce the post-backoff mechanism to improve fairness. When a station transmits successfully, it enters the post-backoff state, where it generates a random backoff number in $[\frac{\beta+1}{2}..\beta]$ for the next upstream frame such that other stations have a higher probability to gain channel access. 
	
In addition, we propose a mechanism for the case when channel access is obtained by node $k$ and there is only one frame to transmit, whose length is smaller than the upstream frame length. In this case, when the contention resolution (at stage $s$) and transmission are finished, node $k$ transmits a special speculative preamble to inform other nodes that the channel is free and the contention resolution is resumed at stage $s$ such that another station is allowed to use the channel.

\section{Analysis}
\label{sec:analysis}
In this section, we analyze the performance of our proposed MAC protocol under non-saturated conditions (stations do not always a packet to send). We analyze the case where stations, denoted by $\mathcal{N}$, send to the BS, whereby a given station $k \in \mathcal{N}$ can transmit using channels $\mathcal{C}_k$, that is, a subset of all channels $\mathcal{C}$. Each channel has a capacity of $\varsigma$ bps. Furthermore, we assume that each station $k$ sends $\lambda_k$ frames per second and each frame has an average payload length of $\bar{L}$ bits. The system is slotted with a slot duration of $\epsilon$, translating into a propagation delay of $\frac{d}{300000}$ for a distance $d$. We let $\beta$ be the maximum backoff value and the maximum stage number corresponds to $s_{max}$. Therefore, each stage takes $(\beta+1+5) \cdot \epsilon$ seconds, whereby $5\epsilon$ is the speculative preamble duration required by the lattice correlation mechanism. The analysis also takes into account two types of probabilistic error: ($i$) bit error occurring with probability $p_{e}^{bit}$, and ($ii$) Lattice correlation error occurring with probability $p_{e}^{corr}$.

The frame duration consists of the upstream and downstream subframes ($T_u$ and $T_d$, respectively):

\begin{equation}
T_f = T_u + T_d.
\end{equation}

We let $q_{k, c}$ be the probability to have a frame waiting for transmission at node $k \in \mathcal{N}$ for channel $c \in \mathcal{C}_k$ at the beginning of a frame:







\begin{equation}
q_{k, c} = \frac{1 - e^{-\delta_{k, c} \cdot T_f}}{P_{succ}(k, c)},
\end{equation}
where $P_{succ}(k, c)$ corresponds to the successful probability to transmit over channel $c$, derived shortly, and $\delta_{k, c}$ is given by:
\begin{equation}
\delta_{k, c} = \frac{\lambda_k}{|\mathcal{C}_k|} \cdot \Bigg\lceil \frac{\bar{L}}{U_{k, c}} \Bigg\rceil.
\end{equation}

$\delta_{k, c}$ gives the traffic rate at node $k$ at channel $c$, whereby station $k$ uses any of its allowed channels ($\mathcal{C}_k$). $U_{k, c}$ corresponds to the average useful bits transmitted during the upstream subframe:
	
\begin{equation}
U_{k, c} = \varsigma \cdot \Bigg (1 - \frac{1}{T_u} \cdot \sum_{s=0}^{s_{max}} \theta_{k, c}(s) \cdot (\beta + 1 + 5)\epsilon\Bigg),
\end{equation}  
where the equation substract the contention duration over all experienced stages $s$. $\theta_{k, c}(s)$, derived shortly, gives the probability to reach stage $s$ given it subsequently successfully transmits.

We let $P_{k, c}(s, j)$ be the probability that station $k$ on channel $c$ selects the contention value (in $[0..\beta]$) $j$ at stage $s$. For state $s=0$ and $j \in [0..\frac{\beta+1}{2}-1]$, we have:

\begin{eqnarray}
P_{k, c}(s, j) & = & \theta_{k, c}(s) \cdot \nonumber\\
	& & \quad (q_{k, c} \cdot (1 - P_{succ}(k, c)) + (1 - q_{k, c})) \cdot \nonumber\\
	& & \quad \frac{1}{\beta+1}, \nonumber\\
\end{eqnarray}
where this case occurs when no post-backoff is done which happens when either there was a frame waiting previously and it was not transmitted successfully or there was not any frame waiting. For $s=0$ and $j \in [\frac{\beta+1}{2}..\beta]$ we have:

\begin{eqnarray}
P_{k, c}(s, j) & = & \theta_{k, c}(s) \cdot \nonumber\\
	& & \quad (q_{k, c} \cdot (1 - P_{succ}(k, c)) + (1 - q_{k, c})) \cdot \nonumber\\
	& & \quad \frac{1}{\beta+1} \nonumber\\
	& & + \theta_{k, c}(s) \cdot q_{k, c} \cdot P_{succ}(k, c) \cdot \frac{1}{(\beta+1)/2},\nonumber\\
\end{eqnarray}
where the additional term is due to the post-backoff mechanism which selects a larger contention value after a successful transmission. $\theta_{k, c}(s)$ is given by:

\begin{equation}
\theta_{k, c}(s)=
\begin{cases}
1, & \text{if $s = 0$},\\
S_{k, c}(s-1), & \text{if $s > 0$}.
\end{cases}
\end{equation}
	For $s > 0$, $S_{k, c}(s)$ gives the probability that a given station $k$ selects the same contention window as another station competing for channel $c$ at stage $s$:
\begin{eqnarray}
S_{k, c}(s) & = & \sum_{j = 0}^{\beta} P_{k, c}(s, j) \cdot \prod_{l} (1 - \sum_{i = 0}^{j - 1} q_{k, c} \cdot P_{l, c}(s, i)) \cdot \nonumber\\
	& & \sum_{l} q_{k, c} \cdot P_{k, c}(s, j).
\end{eqnarray}

	The probability to successfully gain access of channel $c$ at stage $s$ is therefore given by:

\begin{eqnarray}
P_{succ}(k, c, s) & = & \sum_{j = 0}^{\beta} P_{k, c}(s, j) \cdot (1 - p_{e}^{corr}) \cdot \nonumber\\
	& & \prod_{l} \bigg(1 - \sum_{i = 0}^{j} q_{l, c} \cdot P_{l, c}(s, i)\bigg) \cdot \nonumber\\
	& & \quad (1 - p_{e}^{corr}). \nonumber\\
\end{eqnarray}

Considering the probable bit errors and successful probabilities at each given stage, the overall probability to successfully transmit over channel $c$ is given by:

\begin{equation}
P_{succ}(k, c) = (1 - p_e^{bit})^{\varsigma} \cdot \sum_{s}^{s_{max}} P_{succ}(k, c, s).
\end{equation}

\subsection{Delay Analysis}

	We derive the access delay at station $k$ on channel $c$ as follows:

\begin{equation}
\Delta_{k, c}^{ac} = \frac{T_f}{P_{succ}(k, c)}.
\end{equation}

	Averaging the access delays on all allowed channels from station $k$, the average access delay is given by:

\begin{equation}
\Delta_{k}^{ac} = \frac{1}{|\mathcal{C}_k|} \cdot \sum_{c \in \mathcal{C}_k} \Delta_{k, c}^{ac}.
\end{equation}

	Finally, we approximate the overall delay (queuing plus service time) according to an M/M/m (with $m = |\mathcal{C}_k|$) model:

\begin{equation}
\Delta_{k} = \frac{C\bigg(|\mathcal{C}_k|, \frac{\lambda_k}{1 / \Delta_{k}^{ac}}\bigg)}{|\mathcal{C}_k| \cdot (1 / \Delta_k^{ac}) - \lambda_k} + \frac{1}{1/\Delta_k^{ac}},
\end{equation}
where the Erlang C equation $C(m, \rho)$ is given by:
\begin{equation}
C(m, \rho) = \frac{\frac{(m \rho)^m}{m!} \cdot \frac{1}{1 - \rho}}{\sum_{k = 0}^{m-1} \frac{(m\rho)^k}{k!} + \frac{(m \rho)^m}{m!} \cdot \frac{1}{1 - \rho}}.
\end{equation}

	The mean delay is derived as follows:

\begin{equation}
D = \frac{1}{\sum_{k \in [1..N]} \lambda_k} \cdot \sum_{k \in [1..N]} \lambda_k \cdot \Delta_k.
\end{equation}

\subsection{Stability Analysis}

The system is stable if and only if:

\begin{equation}
\lambda_k \cdot \Delta_{k}^{ac} < 1, \forall k \in \mathcal{N}.
\end{equation}

\subsection{Solving the Model}

	The nonlinear system can be solved quickly by means of fixed point iterations. One needs to first fix $q_{k, c}$ and $P_{succ}(k, c)$. Then, iteratively, calculate $P_{k, c}(s, j)$, $P_{succ}(k, c)$, and $q_{k, c}$ until the values do not change significantly (e.g., less than $10^{-8}$). 


\section{Numerical Results}
\label{sec:numericalResults}
	In this section, we compare the throughput-delay performance of our proposed MAC protocol with the standard WiMAX protocol. We use the same configuration, as described in Section \ref{sec:motivation}. The distance is set to $d = 1$ km and each station can transmit over maximally 7 allowed channels ($|\mathcal{C}_k| = 7, \forall k \in \mathcal{N}$). The main parameters are set as follows: $T_f = 0.005$, $\varsigma = 420 \cdot 8$, $\beta = 7$, and $s_{max} = 15$.

\begin{figure}
\centering
\includegraphics[width=.50\textwidth]{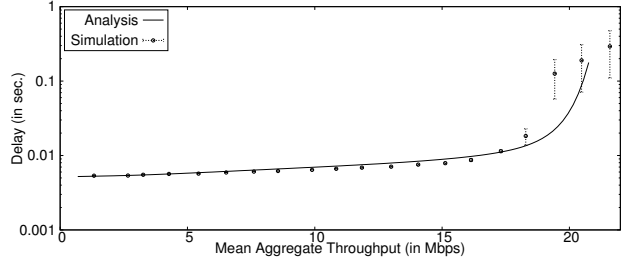}
\caption{Verification of the proposed MAC protocol by comparing analysis and simulation models.}
\label{fig:validation}
\end{figure}

	We first verify the accuracy of the analysis and simulation models of the proposed MAC protocol for $p_{e}^{bit} = 10^{-5}$ and $|\mathcal{N}| = 10$. For ease of verification, we set the payload length to $\bar{L} = 300 \cdot 8$ to avoid fragmentations. We found that our models are accurate, as depicted in Fig. \ref{fig:validation}.

\begin{figure}
\centering
\includegraphics[width=.50\textwidth]{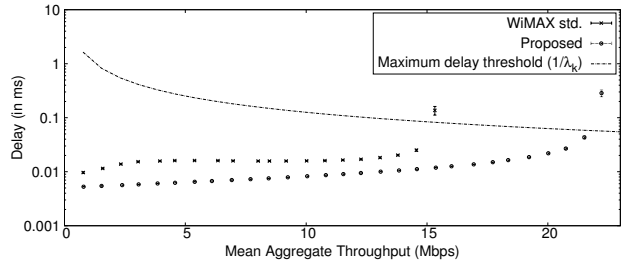}
\caption{Performance comparison of the proposed MAC protocol vs. standard WiMAX ($p_{e}^{bit} = p_{e}^{corr} = 0$, $|\mathcal{N}| = 315$).}
\label{fig:mainRes}
\end{figure}

Next, we set the number of stations to $|\mathcal{N}| = 315$ without any bit errors ($p_{e}^{bit} = 0$) and compare our proposed MAC protocol and WiMAX in Fig. \ref{fig:mainRes}. We set the maximum delay threshold to $\frac{1}{\lambda_k}$ for smart grid monitoring messages, i.e., the delay should be lower than the interval between messages. The maximum aggregate throughput that is able to keep the delay below the maximum delay threshold is found to be 15.2 and 21.5 Mbps for WiMAX and the proposed protocol, respectively. Hence, the proposed protocol is able to improve the maximum throughput by up to $\frac{21.5}{15.2} = 41\%$ for the configurations under consideration. 


\section{Conclusions}
\label{sec:conclusions}
By injecting smart grid traffic traces based on IEC 61850 in our WiMAX simulation model, we observed poor network efficiency performance as the number of nodes increases. We proposed and analyzed a novel MAC protocol for WiMAX by exploiting lattice correlators to quickly solve channel contention and obtain high network efficiency. The obtained results of the proposed MAC protocol performance in terms of throughput and delay have shown ta throughput improvement of up to 41\%.

\bibliographystyle{IEEEtran}
\bibliography{rwpulc}

\begin{thebibliography}{1}
\providecommand{\url}[1]{#1}
\csname url@samestyle\endcsname
\providecommand{\newblock}{\relax}
\providecommand{\bibinfo}[2]{#2}
\providecommand{\BIBentrySTDinterwordspacing}{\spaceskip=0pt\relax}
\providecommand{\BIBentryALTinterwordstretchfactor}{4}
\providecommand{\BIBentryALTinterwordspacing}{\spaceskip=\fontdimen2\font plus
\BIBentryALTinterwordstretchfactor\fontdimen3\font minus
  \fontdimen4\font\relax}
\providecommand{\BIBforeignlanguage}[2]{{%
\expandafter\ifx\csname l@#1\endcsname\relax
\typeout{** WARNING: IEEEtran.bst: No hyphenation pattern has been}%
\typeout{** loaded for the language `#1'. Using the pattern for}%
\typeout{** the default language instead.}%
\else
\language=\csname l@#1\endcsname
\fi
#2}}
\providecommand{\BIBdecl}{\relax}
\BIBdecl

\bibitem{6157575}
{Y. Yan, Y. Qian, H. Sharif, and D. Tipper}, ``{A Survey on Smart Grid
  Communication Infrastructures: Motivations, Requirements and Challenges},''
  \emph{IEEE Communications Surveys \& Tutorials}, vol.~15, no.~1, pp. 5--20,
  Feb. 2013.

\bibitem{lteSG}
{P. Cheng, L. Wang, B. Zhen, and S. Wang}, ``{Feasibility Study of Applying LTE
  to Smart Grid},'' in \emph{Proc., IEEE First International Workshop on Smart
  Grid Modeling and Simulation (SGMS)}, {Brussels, Belgium}, Oct. 2011, pp.
  108--113.

\bibitem{6195796}
X.~Feng, J.~Zhang, Q.~Zhang, and B.~Li, ``Use your frequency wisely: Explore
  frequency domain for channel contention and ack,'' in \emph{INFOCOM, 2012
  Proceedings IEEE}, 2012, pp. 549--557.

\bibitem{wifiNano}
{E. Magistretti, K. K. Chintalapudi, B. Radunovic, and R. Ramjee},
  ``{WiFi-Nano: Reclaiming WiFi Efficiency Through 800 ns Slots},'' in
  \emph{Proc., ACM MobiCom}, {Las Vegas, Nevada, USA}, Sep. 2011.

\bibitem{wcncWiFi}
{M. L\'evesque, M. Maier, F. Aurzada, and M. Reisslein}, ``{Analytical
  Framework for the Capacity and Delay Evaluation of Next-Generation FiWi
  Network Routing Algorithms},'' in \emph{Proc., IEEE Wireless Communications
  and Networking Conference (WCNC)}, {Shanghai, China}, Apr. 2013.

\bibitem{wimaxPerfEval}
{WiMAX Forum}, ``{Mobile WiMAX - Part I: A Technical Overview and Performance
  Evaluation},'' \emph{Technical Report}, Aug. 2006.

\end{thebibliography}
  
\end{document}